\documentclass[prd, twocolumn, nofootinbib, floatfix]{revtex4-1}

\usepackage{amsmath}
\usepackage{graphicx}
\usepackage{dcolumn}
\usepackage{bm}
\usepackage{epsfig}
\usepackage{amssymb,latexsym,mathrsfs}
\usepackage{graphicx}
\usepackage{color}
\usepackage{hyperref}

\hypersetup{
    colorlinks=true,
    linkcolor=red,
    citecolor=blue,
} 

\newcommand{\be}{\begin{equation}}
\newcommand{\ee}{\end{equation}}
\newcommand{\bs}{\begin{split}} 
\newcommand{\bea}{\begin{eqnarray}}
\newcommand{\eea}{\end{eqnarray}}

\newcommand{\op}{\Omega_\phi}
\newcommand{\opo}{\Omega_{\phi,0}}
\newcommand{\wi}{w_\infty}

\newcommand{\al}{\alpha}

\begin{document}

\title{$\Lambda$ is Coming: Parametrizing Freezing Fields} 
\author{Eric V.\ Linder${}^{1,2}$} 
\affiliation{${}^1$Berkeley Center for Cosmological Physics \& Berkeley Lab, 
University of California, Berkeley, CA 94720, USA\\ 
${}^2$Energetic Cosmos Laboratory, Nazarbayev University, Astana, 
Kazakhstan 010000} 

\begin{abstract}
We explore freezing dark energy, where the evolution of the field 
approaches that of a cosmological constant at late times. We propose 
two general, 
two parameter forms to describe the class of freezing field models, 
in analogy to ones for thawing fields, here based on the physics of 
the flow parameter or the calibrated $w$--$w'$ phase space. Observables such 
as distances and Hubble parameters are fit to within 0.1\%, and the dark 
energy equation of state generally to within better than 1\%, of the exact 
numerical solutions. 
\end{abstract} 

\date{\today} 

\maketitle

\section{Introduction} 

Cosmic acceleration of the expansion of the universe can be treated as 
arising from an effective scalar field; this holds even if the physical 
origin is actually, say, from a higher dimensional braneworld. At the 
level of the Friedmann equations, the expansion $a(t)$ can be equivalently 
described by a dark energy equation of state $w(a)$, modulo the matter 
density. 

The physics of an observationally viable universe -- one with a long 
period of matter domination in which density perturbations can grow and 
then a recent period of cosmic acceleration -- group the acceptable 
scalar field behaviors into two classes: thawing fields and freezing 
fields that lie in distinct regions of the $w-w'$ phase space 
\cite{zlatev,caldlin,paths,calde}. Here a 
prime denotes $d/d\ln a$. These classes are related to the competition 
between the driving term of the steepness of the potential and the friction 
of the Hubble expansion. The separation between the classes is enforced 
by avoidance of fine tuning in the field, that the acceleration of the 
field will generically not be so exquisitely balanced that it vanishes. 

Across these classes a highly successful general description is provided 
in terms of two parameters, $w_0$ and $w_a$, corresponding to the present 
dark energy of state parameter and a measure of its time variation. This 
was first derived in terms of exact solutions of the Klein-Gordon equation 
of motion \cite{linprl} and later quantified as providing 0.1\% accuracy on 
observable quantities \cite{calde}. For some purposes, though, one might 
be interested in going beyond observables and seeking an improved 
description of the equation of state $w(a)$ itself. 

An equation of state description for the class of thawing fields has been 
treated in depth in the literature 
\cite{crittenden,algeb,scherrer,dutta,chiba,bond}. Most 
recently, a one parameter form simplifying the general algebraic thawing 
expression of \cite{algeb} has been demonstrated to have 0.3\% accuracy 
in $w(a)$ \cite{1501.01634}. Freezing field models however have been more 
problematic, due to their greater diversity in initial conditions and 
evolution. Here we address the issue of parametrizing the equation of 
state for the class of freezing models. 

Section~\ref{sec:method} discusses the methodology, and we present the 
results in Sec.~\ref{sec:results}, before concluding in Sec.~\ref{sec:concl}.

\section{Methodology} \label{sec:method} 

Freezing fields of interest start in the matter dominated epoch (by which 
is meant either nonrelativistic matter or radiation) with the field 
rolling down a steep potential. At late times (in the future) the field 
freezes as the field enters the shallow region of the potential, and the 
field acts like a cosmological constant, asymptotically approaching a 
de Sitter state (though the field motion may never vanish completely at 
finite times). 
Potentials of interest do not have a nonzero minimum, 
i.e.\ an intrinsic cosmological constant. 

Early freezing models date back to the exponential and inverse power law 
potentials \cite{wett88,ratrap,ferreirajoyce,liddlescherrer} and were 
particularly interesting for having an attractor behavior that brought 
the field from a wide variety of initial conditions onto a tracker 
trajectory at high redshift where the equation of state became constant. 
One problem  
with this was the difficulty of moving the field sufficiently quickly 
off the tracker so that it could attain a sufficiently negative equation 
of state $w\approx-1$ near the present. The potential needed to be 
modified to allow this to happen, e.g.\ within the supergravity inspired 
approach of \cite{braxmartin}. 

An interesting discovery was that higher dimensional braneworld models 
\cite{dgp} and more phenomenological generalizations by adding a term 
$H^\alpha$ to the Friedmann expansion equation \cite{dvalitur} (with 
the DGP braneworld 
corresponding to $\alpha=1$) acted in a very similar way to freezing 
scalar fields, and indeed one could write down an effective potential 
\cite{flow}. 
At early times these behave like an inverse power law potential 
$V\sim\phi^{-n}$ with 
index $n=2\alpha/(2-\alpha)$. 
At late times, such freezers approach $w=-1$ along the phase space trajectory 
$w'=3w(1+w)$, the lower boundary of the freezing region. This corresponds 
to a potential 
\be 
V(\phi)=V_\infty\,\left[1+\frac{3}{8}(\phi_\infty-\phi)^2\right]\ , 
\ee 
where the field asymptotically freezes to the value $\phi_\infty$ 
(while this nominally has a nonzero minimum, it is only an effective 
potential and there is no true cosmological constant, e.g.\ in 
Minkowski space). 

Thus the variety of potentials means that 
there is no expectation that attempting to find a common parametrization 
of freezing field potentials should be successful. Indeed, there are clear 
differences between inverse power law and supergravity inspired potentials. 
Thus, we instead focus on capturing the key physics common to different 
classes of potentials. 

One approach is to follow the physics in terms of the flow parameter 
\cite{flow} 
\be 
F(a)\equiv \frac{1+w(a)}{\Omega_\phi(a)\,(V_\phi/V)^2}\ , 
\ee 
where $\op(a)$ is the fractional dark energy density and $V_\phi=dV/d\phi$. 
The flow parameter was shown to 
be conserved through the matter dominated era, with the value 4/27 for 
the thawing class and 1/3 for the freezing class. During the 
present accelerating epoch $F$ slowly deviates from these asymptotic values 
as the dark energy density increases. The flow parameter is directly 
related to the dark energy phase space evolution and so given a 
parametrization for $F$ one can derive $w(a)$, from 
\be  
w'=-3(1-w^2)\left[1-\frac{1}{\sqrt{3F}}\right]\ . \label{eq:wF} 
\ee 

Another approach to consider is based on the phenomenological 
calibration of the phase space trajectories exhibited in \cite{calde}. 
This showed great success in taking the diversity of equation of state 
behaviors and defining ``stretching parameters'' (leading to particular 
versions of $w_0$ and $w_a$) that focused the models into narrow bands 
or families. One might parametrize the freezing models according to 
band and location along the band. 

We now discuss these two approaches -- the flow and 
calibration approaches -- in more detail.

\subsection{Flow Parametrization} \label{sec:flow} 

We know that the physics of the long, matter dominated epoch forces the 
field evolution into certain paths. This causes the combination of 
dark energy density, equation of state, and steepness of the potential 
to be interrelated in such a way as to keep constant the combination 
in the flow parameter $F$. If we focus on tracker freezing fields, as the 
most attractive due to their insensitivity to initial conditions, then 
the early equation of state is constant, i.e.\ the dark energy density 
$\rho_\phi\sim a^{-3(1+w)}$, and so by Eq.~(\ref{eq:wF}) we see that 
$F=1/3$ during matter domination. 

As the dark energy density grows, $F$ begins to increase from this 
constant, but slowly -- see Fig.~4 of \cite{flow} -- with the deviation 
proportional to the fractional dark energy density $\op$. Since dark 
energy has only become significant within the last e-fold of cosmic 
expansion, and is not fully dominant today, a reasonable ansatz for 
the flow parametrization is 
\be 
F(a)=\frac{1}{3}\,[1-\op(a)]+b\,\op(a)\ , \label{eq:flowop} 
\ee 
where $b$ is a constant parametrizing the specific freezing model. In 
the matter dominated epoch this expression reduces to $F=1/3$ as desired, 
and $F$ again goes to a constant in the future. Since we do not have 
observations in the future, we do not force $b$ to a particular value 
(e.g.\ for all braneworld models it would be $4/3$), but leave it as a 
fitting parameter to observable data. 

Note that $F=1/3$ does not fully characterize the initial conditions as 
this simply leads to $w'=0$ at early times, without determining the 
specific value 
of $w$ in the high redshift tracking regime. This is then an additional 
parameter $w_\infty$, where in the case of early time inverse power law 
potential behavior $w_\infty=-2/(n+2)$. The two parameters of this ansatz are 
then $b$ and either $w_\infty$ or $n$. We will keep the second parameter 
as $\wi$ since this is somewhat more generic. 

To solve the evolution we have a system of coupled equations of motion, 
\bea 
w'&=&-3(1-w^2)\left[1-\frac{1}{\sqrt{3F}}\right]\\ 
\op'&=&-3w\op(1-\op)\ . 
\eea 
One can write a closed form solution for $w(a)$, given by 
\bea 
w&=&\frac{C-1}{C+1}\\ 
C&=&\frac{1+w_i}{1-w_i}\,\left(\frac{a}{a_i}\right)^{-6}\, 
e^{2\sqrt{3}\int_{a_i}^a d\ln a'\,F(a')^{-1/2}}\ . 
\eea 
If one adopts a perturbative expansion about the high redshift value of 
the form $F(a)=F_0+F_1 a^s$, e.g.\ since $\op\sim a^{-3(1+w_\infty)}$ at 
high redshift, one can solve this analytically \cite{09092251} but 
since we are interested in times when the dark energy density becomes 
dominant we will work with the form Eq.~(\ref{eq:flowop}) and the 
numerical solution. 

The flow parametrization has the useful property that it is closely connected 
(much more so than the potential) to what we want: the dark energy equation 
of state $w(a)$.

\subsection{Calibration Parametrization} \label{sec:calib} 

An even more direct approach is to parametrize $w(a)$ explicitly. This 
is of course exactly what the standard $w_0$--$w_a$ form does, but here 
we are focusing specifically on the freezing class and seek increased 
accuracy. One application of $w_0$--$w_a$ did this with an eye on the 
distance-redshift relation, in 
\cite{calde}, where the phase space trajectories of freezing models 
were calibrated through a stretching relation 
\be 
w_a^{\rm calib}\equiv -w'(a_\star)/a_\star\ . 
\ee 
Adopting the value of the temporal stretching parameter $a_\star=0.85$ 
was found to calibrate the freezing field models into tight family bands, 
and the resulting 
\be 
w(a)=w_0+w_a^{\rm calib,d}(1-a) 
\ee 
gave accurate reconstruction of the observable distances and Hubble 
parameters at the 0.1\% level. 

Here we will parametrize the phase space to fit $w(a)$ itself. We will find 
in the next section that this works very similarly to the distance 
calibration, but here $w_a$ is defined by choosing the stretching to 
calibrate the entire evolution $w(a)$. This results in 
the stretching parameter becoming $a_\star=0.82$. Thus this 
approach is also a parametrization with two free parameters, $w_0$ and 
$w_a^{\rm calib,w}$, where the superscript calib,w indicates $w_a$ is fit 
to $w(a)$ rather than to distance or to $w_a=-2w'(z=1)$ or some similar 
constraint.

\section{Results} \label{sec:results} 

To investigate how well the flow parametrization and calibration 
parametrization can reconstruct the true equation of state evolution 
$w(a)$ we consider several typical freezing models. 

The inverse power law (IPL) potential \cite{ratrap,liddlescherrer} 
$V(\phi)=V_\star \phi^{-n}$ tracks at high redshift with a 
dark energy equation of state $w_\infty=-2/(n+2)$. However, it does not 
evolve rapidly toward $w\approx-1$ as the dark energy density grows, 
and so does not accord well with current observations unless $n\ll1$. 

To ameliorate this, the supergravity inspired (SUGRA) potential 
\cite{braxmartin} $V(\phi)=V_\star \phi^{-n} e^{\phi^2/(2M_p^2)}$ enables 
values much closer to $w\approx-1$ to be attained by the present, despite 
acting like IPL at high redshift. For 
the IPL and SUGRA potentials, we solve the Klein-Gordon equation of motion 
numerically (by a fourth order Runge-Kutta technique) to obtain $w(a)$. 

A more phenomenological model of interest is when the dark 
energy contribution to the Friedmann expansion equation takes the form 
of a $H^\alpha$ modification \cite{dvalitur}, where $H$ is the Hubble 
parameter. The case where $\alpha=1$ corresponds to higher dimensional 
braneworld DGP gravity \cite{dgp}; this has the present equation of state 
too far from $w=-1$ to be viable, but $w_0$ approaches $-1$ for smaller 
$\alpha$ (with $\alpha=0$ corresponding to the cosmological constant). 

The $H^\alpha$ model evolution can be solved analytically, with 
\bea 
a&=&\left(\frac{\op}{\opo}\right)^{2/[3(2-\alpha)]} 
\left(\frac{1-\opo}{1-\op}\right)^{1/3}\\ 
w&=&-\left(\frac{2-\al}{2-\al\op}\right)\ , 
\eea 
where a subscript 0 denotes the present. 

\begin{figure}[htbp!]
\includegraphics[width=\columnwidth]{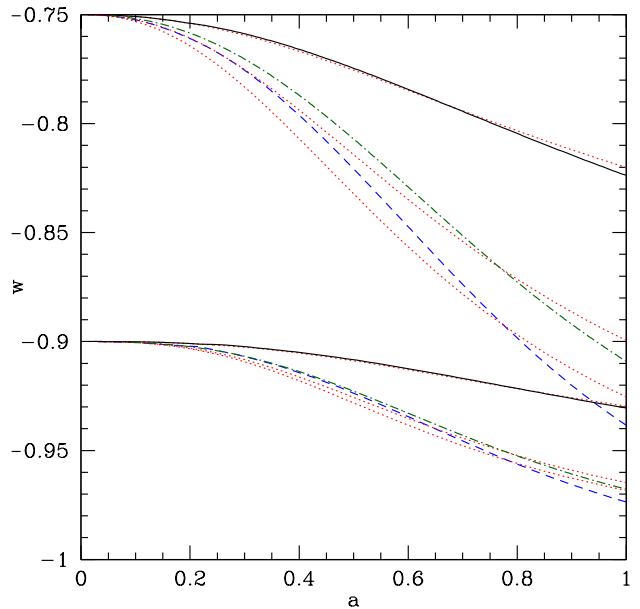} 
\caption{
Flow parametrization (dotted, red) for $w(a)$ compared to exact results for 
IPL (solid, black), SUGRA (dashed, blue), and $H^\alpha$ (dot-dashed, 
green) cases. 
} 
\label{fig:flow3} 
\end{figure}

For each model we will choose values of $w_\infty$ (or equivalently 
$n$ or $\alpha$) and then find the values of the free parameter $b$ from 
Eq.~(\ref{eq:flowop}) that best approximate the exact $w(a)$ functions. 
Figure~\ref{fig:flow3} illustrates that the flow parametrization provides 
excellent approximations, better than 1\% in $w(a)$ for the viable models 
not too far from $w\approx-1$. Even for $w_\infty=-0.75$ cases, $w(a)$ 
is reconstructed to better than 1\% over almost all of cosmic history.

\begin{figure}[htbp!]
\includegraphics[width=\columnwidth]{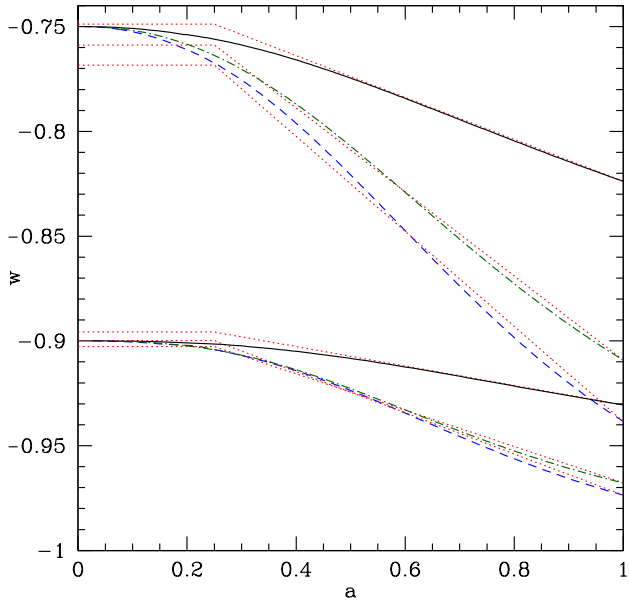} 
\caption{
Calibration parametrization (dotted, red) for $w(a)$ compared to exact 
results for IPL (solid, black), SUGRA (dashed, blue), and $H^\alpha$ 
(dot-dashed, green) cases. 
} 
\label{fig:calib} 
\end{figure}

We repeat the comparison of the exact numerical results with the 
approximation given by the calibration parametrization in 
Fig.~\ref{fig:calib}. In order to reflect the physics of the tracker 
freezing fields at high redshifts, we fix $w(a<0.25)=w(a=0.25)$. Again 
the parametrization is highly successful, accurate at the subpercent level, 
and at the 0.01\%--0.1\% level on the observables of the distances and 
Hubble parameters as a function of redshift.

Tables~\ref{tab:flowacc} and \ref{tab:calibacc} list the maximum deviations 
in $w(a)$, and in the 
observables of distance $d(z)$ and Hubble parameter $H(z)$, over all 
redshifts, for the flow parametrization and the calibration parametrization 
respectively. Deviations in the distance to cosmic microwave background 
last scattering are in all cases near the $10^{-4}$ level. Note that 
freezing models consistent with observations would generally have 
$w_\infty\lesssim-0.9$; indeed the models with $w_\infty=-0.75$ have 
distances to CMB last scattering more than 1\% different from a $\Lambda$ 
model with the same present matter density (and one would have to shift 
the matter density by $\sim0.04$ to get agreement in most cases).

\begin{table}[htbp]
\begin{center}
\begin{tabular*}{0.96\columnwidth} 
{@{\extracolsep{\fill}} l c c c }
\hline
Model & $\delta d/d$ & $\delta H/H$ & $\delta w/w$ \\ 
\hline
IPL ($w_\infty=-0.9$) & 0.03\% & 0.06\% & 0.1\%\\ 
SUGRA ($w_\infty=-0.9$) & 0.04\% & 0.04\% & 0.5\%\\ 
BW ($w_\infty=-0.9$) & 0.02\% & 0.04\% & 0.3\%\\ 
IPL ($w_\infty=-0.75$) & 0.04\% & 0.04\% & 0.4\%\\ 
SUGRA ($w_\infty=-0.75$) & 0.1\% & 0.07\% & 1.4\%\\ 
BW ($w_\infty=-0.75$) & 0.08\% & 0.07\% & 1.1\%\\ 
\end{tabular*}
\caption{Accuracy of flow parametrization 
in fitting the exact distances, Hubble parameters, and dark energy 
equation of state for various dark energy models.  These numbers 
represent the maximum deviation over all redshifts. 
} 
\label{tab:flowacc}
\end{center}
\end{table}

\begin{table}[htbp]
\begin{center}
\begin{tabular*}{0.96\columnwidth} 
{@{\extracolsep{\fill}} l c c c }
\hline
Model & $\delta d/d$ & $\delta H/H$ & $\delta w/w$ \\ 
\hline
IPL ($w_\infty=-0.9$) & 0.01\% & 0.02\% & 0.6\% \\ 
SUGRA ($w_\infty=-0.9$) & 0.03\% & 0.04\% & 0.4\% \\ 
BW ($w_\infty=-0.9$) & 0.03\% & 0.04\% & 0.3\%\\ 
IPL ($w_\infty=-0.75$) & 0.02\% & 0.03\% & 0.9\% ($0.5\%_{z<2}$)\\ 
SUGRA ($w_\infty=-0.75$) & 0.07\% & 0.1\% & 2.4\% ($0.8\%_{z<3}$)\\ 
BW ($w_\infty=-0.75$) & 0.05\% & 0.07\% & 1.2\% ($0.6\%_{z<3}$)\\ 
\end{tabular*}
\caption{Accuracy of calibration parametrization $w_0$-$w_a^{\rm calib,w}$ 
in fitting the exact distances, Hubble parameters, and dark energy 
equation of state for various dark energy models.  These numbers 
represent the maximum deviation over all redshifts, with in some cases 
parenthetical quantities representing deviations in all but the early universe. 
} 
\label{tab:calibacc}
\end{center}
\end{table}

\section{Conclusions} \label{sec:concl} 

Dark energy evolution and the ensuing cosmic acceleration is a competition 
between the steepness of the (effective) potential and the Hubble friction 
of the cosmic expansion. This, together with the long evolution through the 
matter dominated epoch, naturally defines two classes of dark energy: 
thawing models that depart from a frozen, cosmological constant like state 
and freezing models that approach cosmological constant behavior. This 
description holds over a wide range of models for canonical scalar fields 
and modified gravity that 
can be viewed as an effective dark energy. 

A general treatment for both classes is given by the $w_0$--$w_a$ phase 
space parametrization, demonstrated to be accurate in the observables 
to 0.1\%. One can also focus on the dark energy equation of state function 
$w(a)$ itself. While the thawing class has been relatively tractable in 
such treatment, the freezing class has not. Here we investigated 
parametrization of the freezing class at the same accuracy as achieved 
on $w(a)$ for the thawing class. This is more complicated in that the dark 
energy at early times does not have $w=-1$, while at present dark energy 
is not completely dominant. $\Lambda$ is coming, but it is not yet here. 

We demonstrate two approaches to parametrization of $w(a)$ for freezing 
fields, both with reasonably physical levels of motivation. The flow 
parametrization builds on the physics of dark energy evolution during 
the long epoch of matter domination when the flow parameter is constant, 
and parametrizes its deviation as dark energy grows in influence. For 
observationally viable models (not too far from $w=-1$), it achieves 
better than 0.5\% accuracy on $w(a)$ over all cosmic history, as well 
as reconstructing observable distances and Hubble parameters to better 
than 0.06\% accuracy. Even for excursions out to $w=-0.75$ the accuracy 
on $w(a)$ remains near the 1\% level and on the observables to 0.1\%. 

The second approach is the calibration parametrization, using the 
known concept of calibration of the $w$--$w'$ phase space by stretching 
the time variable. This gives the familiar $w_0$--$w_a$ parametrization 
but here $w_a$ is defined by choosing the stretching such that it 
calibrates the entire evolution $w(a)$. Note that the stretching is 
chosen to be model independent, i.e.\ it is fixed for the entire 
freezing class. 
This is successful (with the imposed model independent leveling at high 
redshift) at better than 0.6\% accuracy in $w(a)$ and 0.04\% accuracy 
in the observables for observationally viable models, and remains at 
better than 1\% accuracy on $w(a)$ for excursions out to $w=-0.75$ for 
$z<3$. 

Both approaches use simple, two parameter fits just like $w_0$--$w_a$. 
(Recall that the thawing class actually attained excellent accuracy even 
with a one parameter fit.) For the flow parametrization this is 
$\{w_\infty,b\}$, which provides information on the high redshift, 
tracking state; for the calibration parametrization this is 
$\{w_0,w_a^{\rm calib,w}\}$, very similar to the standard, general 
parametrization. One can easily derive $w_\infty=w_0+0.75\,w_a^{\rm calib,w}$ 
in this approach as well. 

If our universe is such that indeed $\Lambda$ is coming, either of 
these approaches gives a way to characterize accurately, in a fairly 
model independent manner, the dark energy 
equation of state evolution as well as the observables.

\acknowledgments 

I thank Zack Slepian for useful conversations. 
This work is supported in part by the Energetic Cosmos Laboratory and by 
the U.S.\ Department of Energy, Office of Science, Office of High Energy 
Physics, under Award DE-SC-0007867 and contract no.\ DE-AC02-05CH11231.


\end{document}